\documentclass[10pt,=preprint]{aastex}
\usepackage{color}
\begin{document}

\title{Nanolensed Fast Radio Bursts}
\author{David Eichler$^{*}$}
\altaffiltext{*}{Dept. of Physics, Ben-Gurion University, Be'er-Sheba 84105, Israel}
%\altaffiltext{2}{Department of Astrophysical Sciences, Princeton University, Princeton, New Jersey 08544, USA}
%\altaffiltext{3}{Racah Institute of Physics, The Hebrew University, 91904  Jerusalem, Israel}

\begin{abstract}
 It is suggested that fast radio bursts can probe gravitational lensing by clumpy dark matter objects that range in mass from $10^{-3}M_{\odot}$  to  $10^2 M_{\odot}$.  They may provide a more sensitive probe than  observations of lensing of objects in the  Magellanic Clouds, and could find or rule out clumpy dark matter with an extended mass spectrum.

\end{abstract}

\section{Introduction}

Zheng et al (2014)  and Munoz et al. (2016) have suggested that compact dark matter objects  ( MACHOs)\footnote{acronym for massive astrophysical compact halo objects}  of mass M $\sim 30 M_{\odot}$  to $10^2 M_{\odot}$  could be detected by their microlensing of background fast radio bursts (FRBs). Because compact objects in that mass range would  cause time delays between respective arrivals of multiple FRB images of order 1 ms, the same order of the FRB duration itself, the multiple images could be resolved in time.  %If  dark matter is not mainly in the $30 M_{\odot}$ range, 
 Aside from dark matter, the  stars and black holes that are {\it known} to exist in that same mass range constitute only a tiny optical depth 
 ($\ll 10^{-5}$) to lensing  so we should not expect to detect any gravitational echos of FRBs  caused by known objects until hundreds of thousands of FRB have been detected. Even at a detection rate of 20 per day, this would take many decades.  So  $30 M_{\odot}$ MACHOs could stand out above  stars of similar mass even if they comprise only a small fraction of the dark matter. 
 
 On the other hand, the fraction of cosmic mass in familiar stars at $ \lesssim 1  M_{\odot}$ is  of the order  $10^{-2}\rho_c$ - far larger than at $\sim 30 M_{\odot}$.  If the dark matter is in the form of (not-so-massive) compact halo objects (CHOs),  $M\ll 10 M_{\odot}$, then the delay is far less than 1 ms, and would go unnoticed in the inspection of an FRB light curve. However, the constraints  from EROS and MACHO rule out the total being in any one decade in the $10^{-7} M_{\odot} \le M \le M_{\odot}$ mass range, so the only possibility of having $\Omega \gtrsim 0.2$ in CHOs within this mass range is if it is spread out over many decades.  Perhaps this is possible in some rollover scenarios for the early universe (Carr, Kuhnel and Sandstat, 2016). In any case, there is motivation to further search the  $M \lesssim M_{\odot}$ mass range further, to look for lensing manifestations of both known and hypothetical objects in the cosmic density range $10^{-2} $ -  $10^{-1}\rho_c$.

In this letter, it is suggested that lensing delays $t_{delay}$ of much less than 1 ms could be detected by {\it nano}lensing of FRBs. Because FRBs are detected at frequencies of nearly 1 Ghz, lensing delays as small as 1 nanosecond can affect the waveform at the detector. The effect is similar to ``femtolensing'' (Gould, 1992), but is obviously sensitive to a different mass range of the lenses, and has different detection challenges.
It is suggested here than nanolensing of FRB is in principle detectable. At first, this claim might seem surprising because intergalactic dispersion smears out the arrival time of FRBs by a factor of $10^3$ times their true physical duration of order 1 ms. The true pre-dispersion waveform is deduced by measuring the arrival time at different frequencies, which essentially measures the amount of dispersion that was introduced over the travel time of the pulse. However, the dispersion within each frequency bin may be more than the baseband period, and even more than  the lensing delay  $t_{delay}$.   Similarly, as discussed below, delays associated with pulse broadening can exceed that due to the lens. The question arises as to whether enough  information can be recovered in a way that allows detection of lensing delays much shorter than the FRB, and even shorter than the scattering delay.   Though not entirely obvious, the suggestion is made below that in general  it can be.

\section{Long Range Correlation in the Stokes Parameters}
%In the following discussion, we exploit the simplification that the baseband periods $2\pi/\omega$ are typically much smaller than the lensing delay being searched for, which are in turn much shorter than the pulse width $\Delta x \simeq [\int F(x) [x-x_o(t)]^2 dx/\int F(x)dx)^{1/2}/c \sim 10^{-3}$s (where $x_o(t)$ is the center of the pulse at time t), which in turn is less than the spread in arrival times at the detector, typically of the order of 1 s for FRB.   
Suppose the received pulse is of the form 
\begin{equation}
 E(t)=  \int G(\omega') exp[-i\omega' t]d\omega'
\end{equation}
where G is non-negligible over the frequency range $\omega/2\pi \equiv \nu$\ to $\nu + \delta \nu$ of the detector,  which,   for FRB, is of the order of hundreds of Mhz.  For simplicity, we consider only one polarization but define $E$ to be a complex number that contains phase information. We can assume that  G is negligible over frequencies that differ from the detector's frequency range by orders of magnitude, because any other possibility is unnecessary to build a radio pulse at frequencies of hundreds of megahertz that last over far longer timescales $\delta t$ than $\nu^{-1}$; i.e. because $\delta t\gg \nu^{-1}$, it follows that $\delta \nu$ is allowed by the uncertainty relation to be much less than $\nu$, and therefore G need not extend in frequency beyond the intrinsic  spectrum of the source.

 %If the width of $F(x)$ is much larger than 1/k in the region where $G(k)$ is non-negligible, then the Fourier transform of $E(x)$ is very nearly G(k). and the electric field at the detector (defined to be at the origin, while the emission time is defined to be t=0) is given almost exactly  by  \begin{equation}E_d (t) = \int G(k) exp[-i\omega (k) t dk\end{equation}   Now if t is of the order of a Hubble time $10^17.5$s, and $\omega(k) = ck + v_gk^2/2 +\frac{\partial^2\omega,\partial k^2}k^3/6+....$. then the factor $\omega t$ is ------------------------------------------------------------------------------------

 The correlation function $<E^*(t) E(t+\delta)>\equiv\int_{-\infty}^{+\infty}E^*(t) E(t+\delta)dt$ is then given by
 \begin{eqnarray}
E^*(t) E(t+\Delta)  &=& \int_{-\infty}^{+\infty}G^*(\omega')G(\omega'')exp[i\omega' t - i\omega''t -i \omega''\Delta]d\omega'd\omega'' \nonumber \\ 
   &=& \int_{-\infty}^{+\infty}G^*(\omega')G(\omega'')exp[i\omega' t - i\omega''t - i\omega''\Delta]d[\omega'-\omega'']d\omega'' 
\end{eqnarray}
and  
   \begin{equation}
\int E^*(t) E(t+\Delta) dt = 2\pi\int_{-\infty}^{\infty}G^*(\omega')G(\omega')exp[-i\omega' \Delta ]d\omega'.
\label{cor}
\end{equation}

Now if 
\begin{equation}
\Delta\gg\omega'^{-1}
\label{deltaggomega}
\end{equation}
 wherever $G(\omega')$ is nonnegligible,  then $<E^*(t) E(t+\Delta)>$ is exponentially small, whereas if the reverse is true, then the right hand side of equation (\ref{cor}) is approximately $ \int_{-\infty}^{\infty}|G(\omega')|^2 d\omega'$. As a function of $\Delta$,  then,  $\int E^*(t) E(t+\Delta) dt $ is sharply peaked around $\Delta =0$ and this peak has a width of $\sim 1/\omega\equiv1/2\pi \nu$. When $\Delta=0$,

  \begin{equation}
2\pi \int_{-\infty}^{\infty}G^*(\omega')G(\omega')exp[-i\omega' \Delta ]d\omega'  = \int |E^2(t)|dt.
\end{equation}

On the other hand, if $E(t)$ is of the form $a F(t) + b F(t - \Delta_l)$, as it would be for lensed FRBs with $\Delta_l$ the lensing delay between the two images, { then $G(\omega') =   [aG_F + b e^{i\omega' \Delta_l}G_F]$, where $G_F(\omega') = \int_{-\infty}^{\infty} e^{i\omega' t}F(t) dt$,   so $G^*(\omega')G(\omega')=  |a^2 + b^2|G^*_F G_F + 2ab cos({\omega' \Delta_l})G^*_F G_F$.}

 Clearly there is a second peak at $\Delta=\Delta_l \gg 1/\omega$, and
   \begin{equation}
\int E^*(t) E(t+\Delta_l) dt \simeq 2\pi ab \int |F^2(t)|dt
\label{ab}
\end{equation}
Said differently: the short term $\omega'$-periodicity in $G^*(\omega')G(\omega')$,  $  cos(\omega' \Delta_l)$, associated with lensing delays resonates with the $ e^{-i\omega'\Delta}$ factor  in the integrand of equation  (\ref{cor}) when $\Delta=\Delta_l$, whereas, in the absence of lensing, the rapid variation of $ e^{-i\omega'\Delta}$ with $\omega'$ causes self-cancellation of the integrand. 
We therefore note that  another necessary condition for distinguishing lensed sources from unlensed ones is that $G(\omega$)  be sufficiently broadband that the spread in frequencies of the signal, $\delta \omega$, obeys
\begin{equation}
\delta \omega \gg  \Delta_l^{-1}. 
\label{broadband}
\end{equation}
This condition, however, is invariably met by most astrophysical sources (except for line emitters).

{  Note that the right hand side of equation (\ref{cor}) contains no phase information: if the power spectrum $G^*(\omega')G(\omega')$ is known, then this uniquely specifies the left hand side.  Other physical phenomena that cause delay in one light path relative to another, such as scattering, affect the  phase of each Fourier component $G(\omega')$,  but their effect on the correlation function $ <E^*(t)E(t+\Delta)>$ is expressed in equation (\ref{cor}) through  the rapid fluctuation of $G^*(\omega')G(\omega')$ } 
%Note that if the lensing delay is greater than  the baseband periods {$2\pi/\omega$} of the FRB, 

{ Also note a point that will prove to be important below, when diffractive scintillation is discussed:   although the $e^{i\omega'\Delta}$ causes near cancellation of the integral  (\ref{cor}) when $\Delta \gg1/\omega$, rapid  fluctuation in $G^*(\omega')G(\omega')$ with $\omega'$, as is produced by scintillation, does {\it not} cause cancellation, even when the fluctuation scale $\delta \omega'$ is much smaller than $1/\Delta_l$ (i.e. the scattering delay is much larger than $\Delta_l$),  because  $G^*(\omega')G(\omega')$ is positive definite.}
 
The above assumes that the lensing delay $\Delta_l$ is well defined to better than $1/\omega$ for values of $\omega$ with a significant spectral component. If $\Delta_l$  is by hypothesis less than the duration of the undispersed pulse, $\tau$, which is of order 1 ms, then the spread in the lensing delays  is only  of order $(R_s/R_E)^2\Delta_l < (R_s/R_E)^2\tau$, where $R_s$ and $R_E$ are respectively the apparent source radius and the Einstein radius of the lens. [That the delay is only second order in $(R_s/R_E)$ follows from Fermat's principle, namely that the propagation time is a local minimum relative to all other neighboring trajectories, e.g. (Narayan and Bartleman 1996).] If FRBs are at cosmological distances, then, as the apparent source size $R_S$ is limited to the  FRB  pulse width ($\sim 10^{-3}c$ s), and  $R_E^2$ is of order $c^2H^{-1} \Delta_l $, it follows that  the spread in delays, which is of order $(R_s/R_E)^2\Delta_l \sim R_s^2/c^2H^{-1} $,  is indeed much less than 1 ns.

Similarly, the difference in the dispersion time $t_d \sim 1 $ along the two light paths, if the intergalactic medium is smooth,  of the order $\delta \theta t_d$, where $\delta \theta$, the angular separation of the images, obeys $\delta \theta ^2/H \sim 2 \Delta_l$. If $\Delta_l \ll 1$ ms, then $\delta \theta \lesssim 10^{-11}$. Conceivably, the intervening  medium is clumpy on a much smaller scale than $c/H$, so that the difference  in dispersions - in this case governed by the sharp density gradients of the clumps -   is larger. This would not destroy the effect;  the delay is nonetheless well defined.  It would be due to dispersion differences rather than the lens, but the existence of the multiple paths - hence the existence of the lens - would be demonstrated.

A related question is whether the delay is constant over the course of the FRB. While the apparent source size of an FRB is only about 1 milli-lightsecond, the bulk Lorentz factor $\Gamma$  of the emitting region must be at least of order $10^{3.5}$ (Lyubarsky, 2014), so the emitting region may move relative to the observer with an apparent velocity of $\gtrsim 10^{3.5}c$.  
Now for a source at a cosmological distance $c/H$, the Einstein radius of a lens of mass m is $ [(c/H)(2Gm/c^2)]^{1/2}\simeq [10^{16}(m/M_{\odot})]^{1/2} \rm cm$. So the apparent time to cross the Einstein radius is $\sim 10^2[10^{3.5}/\Gamma] [m/M_{\odot}]^{1/2}$ s, and, over the 1 ms duration, the change in the source position could be at most $10^{-5}[10^{3.5}/\Gamma] [m/M_{\odot}]^{-1/2} R_E$. The change in the delay $\Delta_l$ due to the motion of the emitting region  would then be less than $10^{-10}[10^{3.5}/\Gamma]^{-2}[ [m/M_{\odot}]^{-1}\Delta_l \simeq 10^{-15}[\Gamma/10^{3.5}]^2 $ s, which, for imaginable values of $\Gamma$, is much less than the base band period.

The importance of condition (\ref{cor}) over the full frequency  range of significant $G(\omega)$, together with the broadband requirement (\ref{broadband}) on G is emphasized. The effect suggested here is not apparent in any formalism that assumes a monochromatic wave because adding a monochromatic  wave to a delayed version of itself merely shifts the phase, and  changes the amplitude by a modest factor (e.g. Debuchi and Watson (1986), Stanek,  Paczynski and Goodman, 1993). This is insufficient to reveal to the observer whether a delay has occurred or not. 

 The diffraction width over a Hubble distance $c/H$  is  the diffraction angle $\delta \theta \simeq \lambda/R_E$ times $c/H$. This is small relative to the deflection length, which is of the order $R_E\simeq (2Gm/cH)^{1/2}$ if 
\begin{equation}
\lambda \ll 2Gm/c^2,
\end{equation}
and  for $\lambda \sim 10^2$ cm,  this imposes the condition $m\gg 10^{-3} M_{\odot}$. If this condition is not met, diffraction overwhelms the effect of the lens and there are no lensing effects to speak of.

The CHIME array is expected to detect $\sim 20$   FRB per day  (Kaspi, private communication), hence a sample set of many thousand will be collected over several years.  Self-correlation can pick out multiple images even with large brightness contrasts so the nanolensing cross section for any distant source is $\pi R_E^2 $, where $R_E$ is the Einstein radius. The optical depth for nanolensing of FRBs at high redshift   is of order $\Omega_l$ where $\Omega_l$ is the fraction of the critical density that is in the form of compact objects  that play the role of the lens. So for sample sets exceeding $10^3$ in number, the method can be checked with known objects such as stars. The method in principle can limit $\Omega_M$ to $\lesssim 10^{-2}$ in the  sub-$M_{\odot}$ mass range. This would be enough to rule out (or confirm) the possibility the dark matter is in clumps even with an extended mass spectrum.

Fast Fourier Transform radio telescope arrays (Tegmark and Zaldarriaga, 2008) typically correlate signals over the different antennae and thus establish the direction  of any source. Clearly the phase information at each antenna must be accurate to 
a time resolution of much better than $\nu^{-1}$ and stored long enough to compute the correlation. For any direction (other than the zenith) a given signal appears delayed in some antennae relative to others.  A lensed FRB, on the other hand, creates a long term time correlation within any given single antenna. It is important that the  design of the telescope include this possibility - i.e. that phase information be stored long enough\footnote{or at least buffered long enough to determine whether an FRB has occurred, and then, if so, to store the data with full phase information for further analysis} and then searched for self time-correlation within individual antennae. 

\section{ The Effects of an Inhomogeneous Plasma  on the Signal}

FRB are strongly scattered (Masui et. al, 2016) and this scattering can cause arrival delays  of order 1 ms, which is by hypothesis larger than the gravitational lens delay.  It should be emphasized that equation (1), which refers to the {\it received } pulse at the detector, does not deny  this possibility.  The signal, until it arrives at the detector, can experience whatever processing one would care to imagine by the medium through which it propagates, including delays larger than the delay associated with the gravitational lens. What is assumed in the later analysis is simply that whatever delays are induced by the medium are the {\it same} for both of the gravitationally lensed images following the lens. Careful consideration suggests that this can be  the case.

Observationally, interstellar refractive scintillation changes the observed intensity over a timescale of hours, {diffractive scintillation of pulsars over a time scale of minutes}, and interplanetary scintillation over a timescale of seconds. This is not fast enough to influence one gravitationally lensed image relative to another in our context, where the delay is much less than 1 ms. Note that intensity changes are caused by the relative changes in phase over different nearby paths. The spatial separation of  gravitationally lensed images would be much less than for multiple images associated with scintillation, hence the timescale for change of relative phase would be slower than for scintillation.  Moreover, the following argument suggests that even the {\it absolute} rate of change of phase by interstellar turbulence is too slow to matter here. 

The absolute change in phase due to a medium of fluctuating electron density $<N^2>^{1/2}$ with a cell size of $a$ and a thickness $ l $ is, {for a given ray path} 

\begin{equation}
\phi = \lambda r_e a [l/a]^{1/2}<N^2>^{1/2}
\end{equation}
where $r_e$ is the classical radius of the electron (Alurkar, 1997).  So the absolute rate of change of the phase is 

\begin{equation}
d\phi/dt = \lambda r_e a [l/a]^{1/2}<N>[d ln N/dt]
\end{equation}
where $dln N/dt$ can be written as $\nabla \cdot v  $.  For multiscale turbulence, each scale a has an associated velocity $v(a)$, and $\nabla \cdot v = v(a) \epsilon/a$ 
where $\epsilon$ is the compressibility factor.  For a sound wave $\epsilon =1$ whereas for a non-compressive motion, such as pure shear, $\epsilon = 0$. For Kolmogorov turbulence, $v(a) \propto a ^{1/3}$  so $a [l/a]^{1/2}<N>[d ln N/dt] \propto a^{1/6} \epsilon$. But $\epsilon$ scales as $v^2/c_s^2$, where $c_s$ is the sound velocity, and this is $\propto a^{2/3}$. So the quantity $a [l/a]^{1/2}<N>[d ln N/dt] \propto a^{1/2} $ decreases with a, and it is the largest scales that dominate the absolute rate of change of phase.

In the interstellar medium we can take $l$ to be 1 Kpc, and a to be maybe 100 pc. N is at most $1/cm^3$, $\epsilon \le1$, and $v(100 pc) \lesssim 10^{6.5}$ cm s$^{-1}$. So for $\lambda = 30 $cm, 
\begin{equation}
d\phi/dt = \lambda r_e a [l/a]^{1/2}<N>[d ln N/dt] \lesssim 10^{-10.5+6.5}s^{-1}\simeq 10^{-4} s^{-1}
\end{equation}

This gives a time scale of  at least $10^4$s for an absolute change of phase, i.e. several hours,  similar to the observed timescale of variation of BL Lac objects, which is attributed  to  refractive scintillation.\footnote{ When there is diffractive scintillation, the number $\cal N$ of independent cells of size a that diffract  light into the detector is proportional to $(a/\lambda)^{-2}  a^{-2}$ where $a/\lambda$ is the width of the diffraction peak, and the total rate of phase change for $\cal N$ randomly changing components may be ${\cal N}^{1/2}$ times that of a single component, but this is still not enough to cause phase changes over  1 ms or less.} Note that the {\it relative} change of phase  between two nearby paths might be much smaller than the absolute change of either, but could not be much greater. 

The motion of the solar system  and Earth is of the order of $v_E =30$ km/s, and the rate of change of phase for a given scale $a$ in a slab of thickness $l$ due to this motion is 
\begin{equation}
d\phi/dt = \lambda r_e a [l/a]^{1/2}v_E \cdot \nabla N \sim \lambda r_e [l/a]^{1/2}\delta N(a)
%\simeq 10^{-10.5+6.5}s^{-1}\simeq 10^{-4} s^{-1}
\end{equation}
where $\delta N (a) = a \nabla N$.
While the factor $[l/a]^{1/2}$ can increase with decreasing $a$ as fast as $a^{-1/2}$, the amplitude of density fluctuations, as argued above,  probably decreases as least as  fast as $a^{2/3}$, so even here $d\phi/dt$ is expected to decrease with decreasing  a. So purely temporal separation   of the two gravitationally lensed images would not lead to a significant phase change of one relative to the other.  {The angular separation, in the case of diffractive scintillation, is discussed below.} %Moreover, Masui et al (2016) report normal interstellar scintillation  of FRBs - nothing out of the ordinary.

  {The above discussion has assumed that each image of the gravitational lens follows a well defined path to the observer. If the source is small enough to undergo diffractive scintillation, this is not the case because each image is in fact the contribution of many different paths. The smallest scale density fluctuations, which have the broadest diffraction peaks ($\propto \lambda/a$), contribute in greater numbers ($\propto a^{-4}$)  to the signal at the detector, so even if the amplitude of the density fluctuations decreases as $a^{2/3}$,  the larger number of cells contributing makes the smaller $a$ dominate the overall phase change in time. 
Empirically, high latitude pulsars within our Galaxy scintillate over a time scale of minutes, say $10^2$ s,  and this is generally interpreted as the time scale over which the pulsar moves across one  cell in the turbulence  and on to the next (e.g. Narayan, 1992)  so that its line of sight now passes through the latter.  Assuming that the pulsar moves across the sky at a speed of order $2 \cdot 10^7$ km/s, the smallest scale that causes significant phase change is then of order $2 \cdot  10^9$ cm.  This is  a very rough estimate, and it provides a rough estimate of the angular distance, $2 \cdot 10^9/d$ cm in the observer's sky,  over which the brightness change induced by scintillation is correlated.  Here $d$ is the thickness of the interstellar disk $\sim 10^{20.5}$ cm.

A remaining question is whether the scattering of the FRB in the host galaxy raises the effective transverse size of the source, y, enough to cause significant dispersion in the gravitational delay.  The scattering delay for FRB 110523 at 800 Mhz is $\delta t =1.66$ ms (Masui etl al, 2016), and also an implied delay, given a scintillation decorrelation band width of $\sim 1 $Mhz, a delay of $ \sim 1 \mu$s, which is typical of the scattering delay in our own Galaxy.  The delay $\delta t$ due to scattering is given by 
\begin{equation}
\delta t = (y/2s^2) (l/s)^{1/2}
   \end{equation}
   where $s$ is the scattering length { and $y$ is the perpendicular distance between the scattering site and the line of sight} .
      The largest value of y for a given delay is obtained assuming that there is a single scattering ($l=s$ ) over the path length out of the disk of the host galaxy. Taking this length to be $l=\ 163 \rm pc$, which is a typical path length through a spiral galactic disk at a typical angle, the light travel time is $l/c =1.66 \cdot 10^{10}$s,  so the scattering angle due to the $\mu$s delay would be $\theta\simeq y/s = (2 c\delta t/s)^{1/2}  = 1.5 \cdot 10^{-8}$  whence $y\lesssim 10^{13}$ cm.      
      The  1.66 ms delay is much larger than the usual scattering delay for propagation through a galactic disk, and, as argued by Katz (2016), this suggests that the scattering occurs much closer to the source than $l$. Denote this separation  $\eta l; \, \eta \ll 1$.  The implied scattering angle is then $1.4 \cdot  10^{-6.5}\eta ^{-1/2}$, and $ y \simeq 2.3 \cdot 10^{14} \eta^{1/2} $ cm. Assuming the FRB is at a distance of 1 Gpc,  the Einstein radius for a lens of mass $m_l$ is $R_E = 10^{16.25}(m_l/M_{\odot})^{1/2}$ cm. %For $\ eta\le 0.1$    
       The dispersion in the gravitational lens delay is $[{{y}\over{R_E}}]^2/2 \lesssim  10^{-4.}\eta(m_l/M_{\odot})^{-1} $, so, as the gravitational  delay is  of the order of $10^{-5} (m_l/M_{\odot})$ s, its dispersion  is then much less than 1 ns. 
      
      {{ \it Scattering Delay}: Finally, it could be asked whether gravitational delays, which occur in a very small minority, $\lesssim 10^{-2}$,  of cases, could be detected above the scattering delay caused by density fluctuations in the interstellar plasma  which, at high Galactic latitude, are typically 1 $\mu$s at Ghz frequency (Cordes and Rickett, 1998).    Such scattering delay $\Delta_s$ would thus be larger than the gravitational  lensing delay $\Delta_l$ when the gravitational lens mass is less than $0.1 M_{\odot}$.  As the scattering delay is distributed over a wide range of values $\{\Delta_s\}$ (comparable in width to the mean  $\Delta_s$), the arrival time of each image, in the case of a gravitational lens, is accordingly spread over a larger range of $\Delta$ than their separation $\Delta_l$, and the question is whether this would make them difficult to distinguish (as one might suppose) from the single peak one would obtain in the absence of a gravitational lens. To address this question,  first note the ordering of time scales in the problem: $ 2\pi \nu \gg  \omega'_c\gg 1/\tau \gg 1/t_{sc}$, where $2\pi \nu$ is the characteristic  carrier frequency of the FRB at which the FRB is detected, $\omega_c \sim 1 \rm Mhz$ ($\Delta_s \sim 10^{-6}$s)  is the frequency   interval over which the scintillation pattern decorrelates (typical delay in travel time associated with the interstellar diffraction), $\tau \sim 10^{-3}$s is the duration of the FRB, and $t_{sc} \gtrsim 10^2$s is the diffractive scintillation time scale for point sources due to Galactic density fluctuations.   In the absence of gravitational lens delays, the spectrum of a scintillating source can be written as 
      \begin{equation}
      G_F(\omega') = \int \int g (\phi_x,\phi_y,\omega') e^{i \phi(\theta_x,\theta_y,\omega')}d\theta_x d\theta_y  
      \end{equation}
      where $\theta_x $ and $\theta_y$ are the two local sky coordinates in the vicinity of the source direction, and $g (\phi_x,\phi_y,\omega')$ is the spectrum before passing through the scattering screen weighted by the probability that light is scattered back into the line of sight from this direction.   When there is a gravitationally lensed signal, the delayed image can be written as 
        \begin{equation}
      G_d(\omega') =   \int \int g_d(\phi_x,\phi_y,\omega') e^{i  \phi_d(\theta_x,\theta_y,\omega')}d\theta_x d\theta_y 
      \end{equation}
      Here $\phi_d(\theta_x,\theta_y,\omega')$ is the phase of the delayed image in a given pixel. Since the scattering screen is assumed to be stationary, and the scattering angle is very small,  $\phi_d(\theta_x,\theta_y,\omega')$ can be taken to be equal to $\phi_(\theta_x,\theta_y,\omega')$, whereas the function $g$ changes with the angle of incidence on the scattering screen. The total spectrum can then be written as

      \begin{equation}
      G(\omega') = \int \int   \left[
 (ag e^{i \phi(\theta_x,\theta_y,\omega')} +  bg_d e^{i\phi(\theta_x,\theta_y,\omega')}  exp[i\omega'\Delta_l ] ) \right] d\theta_x d\theta_y  
      \end{equation}
      
      \begin{equation}=   \int \int \left[(ag e^{i   \phi(\theta_x,\theta_y,\omega')} +  bg e^{i \phi(\theta_x,\theta_y,\omega')}  exp[i\omega'\Delta_l ] )                          %\\[ + b g_d (\phi_x,\phi_y,\omega') e^{-i\omega'  \phi(\theta_x,\theta_y,\omega')]exp[i\omega'\Delta_l ]    - b g (\theta_x,\theta_y,\omega') 
     +br(\phi_x,\phi_y,\omega') e^{i  \phi(\theta_x,\theta_y,\omega')}exp[i\omega'\Delta_l ]   \right]     d\theta_x d\theta_y 
     \label{G()}
     \end{equation}
     where $r(\phi_x,\phi_y,\omega')\equiv
     g_d(\phi_x,\phi_y,\omega') - g(\phi_x,\phi_y,\omega') $, and finally 
     \begin{eqnarray}  
         G(\omega') 
         %&=& \int \left[   [a+bexp[i\omega'\Delta_l ]\int \int g e^{i\phi(\theta_x,\theta_y,\omega')} d\theta_x d\theta_y 
     %+bexp[i\omega'\Delta_l ]\int\int r(\phi_x,\phi_y,\omega') e^{i  \phi(\theta_x,\theta_y,\omega')}  d\theta_x d\theta_y \right] e^{i\omega' t}dt \\
     &=&  (a+bexp[i\omega'\Delta_l] ) G_F  + bexp[i\omega'\Delta_l ] R .
           \end{eqnarray}
        where    $R(\omega') \equiv   \int\int r(\phi_x,\phi_y,\omega') e^{i  \phi(\theta_x,\theta_y,\omega')}d\theta_x d\theta_y .$
      and 
      \begin{equation}
      G^*G = (a^2 +b^2)G^*_F G_F + 2ab G^*_F G_F cos(\omega'\Delta_l) +b^2(G_F^*R +R^*G_F) +b^2 R^*R+ ab(G_F^*Re^{i\omega' \Delta_l} +R^*G_F  e^{-i\omega'\Delta_l})    
      \label{G*G}
      \end{equation}
    
       When $R\ll G_F$,  the last three terms of equation (\ref{G*G}) can be neglected. As each pixel on the sky sees the same gravitational lens delay,   the $[a+bexp[i\omega'\Delta_l ]$ factor can be be taken out of the integral over $d\theta_x d\theta_y$ in equation (\ref{G()}), so  $G(\omega')$ assumes the same form as before.      As discussed above, the fact that integrating over different light paths in the $\theta_x,\theta_y$ plane introduces rapid decorrelation  in $G^*(\omega')G(\omega')$ with changing $\omega'$  - the frequency decorrelation scale associated with the scintillation  ($\sim 1$ Mhz for the Galactic disk) -  does not greatly diminish the long term correlation produced by the gravitational lens.  It is true that the fluctuations in    $G^*(\omega')G^*(\omega')$ may introduce  Fourier components of this function at scales $\Delta \sim \Delta_s\gg 1/2\pi \nu$, leading to some long term correlations at $\Delta \sim \Delta_s$ but these correlations, should they exist, would be much weaker than that  produced by the gravitational lens, because only a tiny fraction of the signal is subject to any particular value within the wide range of scattering delays $\Delta_s$, in contrast to the gravitational lens delay,  to which the entire second image is subjected.   
  This is seen by writing $I(\omega')\equiv G^*(\omega')G(\omega')\equiv <I> + \delta I(\omega')   $, where $<I>\equiv \int E^*(t')E(t') dt'$,\footnote{Here the time interval over which the integral is performed is long compared to the scattering delay but short compared to the duration of the burst so that the time average of the intensity can be meaningfully defined. Similarly, the scattering delays are assumed to be long compared to the characteristic frequency $2\pi \nu$ of the carrier wave so that an interval in $\omega'$ can be defined over which $<I> $ is meaningfully described as a function of $\omega'$ and $\delta I(/\omega')$ can average out to zero over this interval.} and  $<\delta I> = 0$.  We can then write the Fourier transform variable\footnote{We write the Fourier transform variable as $\Delta$ in contrast to observer time t because the propagation delays implied by the frequency scale of variation of $\delta I(\omega')$ are much smaller than the timescale of the scintillation itself.} of I as 
        \begin{equation}
       { \cal I}(\Delta) \equiv \int_{-\infty}^{\infty} I(\omega') e^{i\omega'\Delta} d\omega'.= <I>\delta(\Delta) +\int_{-\infty}^{\infty}  \delta I(\omega' ) e^{i\omega'\Delta} d\omega'.
        \end{equation}
        It is convenient at this point to consider a finite frequency interval  $[\omega_1, \omega'_2 ]$ over which the source spectrum is more or less constant, and consider discrete Fourier modes   of the variable $I(\omega')=\sum_{i} a_i cos(\omega' \Delta_i) +b_i sin(\omega' \Delta_i)$,\footnote{for simplicity I henceforth neglect the $sin(\omega' \Delta_i)$ components} where the values of $\Delta_i$ are separated by $2\pi/[\omega'_2 -\omega'_1]$. The larger frequency $\omega'_2$ can be taken to be of order the carrier frequency $2\pi \nu$.  Then $a_0  = <I>$ and $\delta I(\omega') = \sum_{i\neq0} a_i cos(\omega'\Delta_i)$
        Assuming the square of the mean  intensity of a scintillating object $<I>^2=a_0^2$
        %[\int_{-\epsilon}^{+\epsilon}{\cal I}d\Delta]^2$, 
        %where $\epsilon $ is slightly larger than the true width of the delta-function, 
         is comparable to the variance  $\sum_{i\neq0} a_i^2$, it follows that $a_0$ exceeds  $a_i|_{i\neq0}$ by roughly the factor $N^{1/2}$, where N is the number of modes that have significantly non-zero values for $a_i$. As N is of order $\Delta_s[\omega'_2-\omega'_1]/2\pi$ where the delay $\Delta_s$ is of order the reciprocal of the the scale of variation of $\delta I$ with frequency, it follows that $N\gg 1$ and that $a_0 \gg a_i|_{i\neq0}$.
         % $[\int \delta I^2(\omega') d\omega' ] =[\int_{|\Delta|\ge \epsilon} {\cal I}^2(\Delta)d\Delta] $, then   the area under the delta-function term $<I>\delta(\Delta)$ is comparable to the area under $[\int_{\Delta\neq 0}   {\cal I}^2(\Delta) d\Delta] \sim {\cal I}(\Delta \sim t_d) t_d^{1/2}$ where ${\cal I}(t_d)$ is the value of $\cal I$ at typical delays $t_d$. Given that the  true width of the delta-function is of order $1/T$ where T is the duration of the FRB,  and the width of ${\cal I}(\Delta) $ is of order the delay $t_d$ introduced by the scattering through intervening plasma fluctuations, it follows that the height of the delta-function peak is higher than that of the background delays $t_d$ by a factor of the order of  $T/t_d \gg 1$.
      It then follows that the the correlation caused by the gravitational lens would easily stand out, even if somehow weakened considerably, above the spectrum of delays caused by the density fluctuations in the intervening plasma. 
      
      Now suppose that R cannot be neglected because the delayed image makes an angle with the original image that is comparable to or larger than the angular decorrelation scale of the scintillation pattern. Then the term  $ab(G_F^*Re^{i\omega' \Delta_l} +R^*G_F   e^{-i\omega'\Delta_l})$ must be comparable to  $2ab G^*_F G_F cos(\omega'\Delta_l)  $ and in fact, the ensemble average of      $ab(G_F^*Re^{i\omega' \Delta_l} +R^*G_F e^{-i\omega'\Delta_l})$  is given by 
      
      \begin{equation}
         < ab(G_F^*Re^{i\omega' \Delta_l} +R^*G_F)  e^{-i\omega'\Delta_l})> = 2ab \eta(\omega') < G^*_F G_F >cos(\omega'\Delta_l)
      \end{equation}
      where $\eta(\omega') \equiv (-1+<G^*(\omega')G_d(\omega')>/<G^*_F G_F)>)$. Thus if $G_d$ is completely uncorrelated with $G$, then $\eta=-1$ and the fifth term on the right hand side of  equation (\ref{G*G}) exactly cancels the second term, so there is no coherent interference between the two images. 
     For any particular event, of course, the value of  R is generally different from -1 and equation (\ref{G*G}) has a component proportional to $e^{1\omega'\Delta_l}$ of order $abG^*G$, but  the integration of $G^*(\omega')G(\omega')$ over a range of $\omega'$ that is large compared to the correlation frequency interval $\omega'_c$, may be nearly equivalent to an ensemble average.   In any case,  if there is even a small residual correlation,  then the  correlation $<E^*(t)E(t+\Delta_l)>$, albeit weakened, remains non-zero, and is possibly detectable. Thus the marginal case where the angle between the images just happens to be  not too much larger that the angular decorrelation scale of $G(\omega)$ might still allow for a detection of the time correlation of the Stokes parameters of the FRB signal over $\Delta_l$. 
      
      The angle $\theta_c = t_{sc} \cdot d\theta/dt $ over which scintillation patterns are decorrelated by 1/e in the Galactic disk can be determined by measuring both the proper angular motion $d\theta/dt$ of the pulsar and the correlation timescale $t_{sc}$  for the scintillation pattern.  For the high latitude pulsars 1115+5030, 1239+2453, and 1509+5531, $\theta_c$ is determined in this way to be $5\cdot 10^{-12}$, $8 \cdot 10^{-12}$, and $2\cdot 10^{-12}$, respectively (Cordes and Rickert,  1998). This is comparable to the angular separation between the two images of a source at distance 1 Gpc,  lensed by an object  of a solar mass, so for lenses much more massive than this, the correlation over $\Delta_l$ would be weakened. More distant sources allow somewhat higher lens masses.
   
          %$\cal{I}\sim \int_ $
      
      %In this regard it is important to note that gravitational delays are achromatic whereas scattering delays are strongly frequency dependent, being proportional to $\nu^{-4.4}$.  At somewhat higher frequencies, say several Ghz, this delay would be much smaller. Now, even if the FRB is detected by a detector at lower frequency, the fact remains that if the source emits significant power at $\nu \gtrsim 2 $Ghz,  this power contributes to the right hand side of equation (\ref{cor}),  and therefore to the left hand side.  So if the left hand side can be measured, even at $\Delta \gg 1/\nu$, it should in the case of a gravitational lens delay $\Delta_l$ show a sharp peak  at $\Delta=\Delta_l$, if FRB power extends into the multiGhz frequency realm, provided that the detector apparatus does not filter out the multiGhz  signal and measures the true, instantaneous { $E(t)$}. }}
      
\section{Summary}
We have suggested that nanolensing of FRBs - i.e. gravitational lensing delays of $10^{-8}\le \Delta_l \le 10^{-3}$s  -  can probe the distribution of mass in compact objects in the individual mass range $10^{-3} - 10^2 M_{\odot}$  with unprecedented sensitivity.  The technique could not only set limits on dark matter, but apparently investigate the mass distribution of brown dwarfs if it is not exceeded by that of dark matter.  { The angular image separation  $10^{-11.5}(m/M_{\odot})^{1/2}$ due to the gravitational lens may, for $m\gtrsim M_{\odot}$, be comparable to the correlation angle of the scintillation pattern of a point source  due to small scale plasma density fluctuations in our  Galaxy.  Because of the phase decorrelation of the two images in this situation, the coherence of Stokes parameters over the gravitational lens delay,  $\Delta_l$, is diminished, We have argued, however, that even at $m\sim M_{\odot}$, this is not necessarily lethal.

{  Leaving aside the history of ideas about lensing, the technique  proposed here has nothing to do with FRB {\it per se}. Any radio source at cosmological distances would work as well, provided it is compact enough that its gravitational lens delay is well defined to  better than the period of the carrier wave $2\pi/\omega$.}

I thank Drs. G. Molodij, R. D. Blandford, J. Goodman,    Y. Lyubarsky, A. Zitrin, V. Kaspi , A. Spitkovsky,  J. Silk, N. Bahcall, S. Phinney, J. Cordes, B. Zackay, and the referee for helpful conversations.  This research was supported by an ISF-UCG grant, by the Israel-US Binational Science Foundation, and by the Joan and Robert Arnow Chair of Theoretical Astrophysics.

\end{document}